\documentclass[a4paper,11pt,dvipsnames]{article}
\pdfoutput=1 %
\usepackage{jheppub} %
\usepackage{epsfig}
\usepackage{amsfonts,amsmath,amssymb}
\usepackage{graphicx}  
\usepackage{dcolumn}   
\usepackage{bm}        
\usepackage{slashed}
\usepackage[export]{adjustbox}
\usepackage[caption=false]{subfig}
\usepackage{hyperref}
\newcommand{\mueee}{\mu^- \to e^- \, (e^+ e^-) \, \nu_\mu \bar{\nu}_e}
\newcommand{\mysmall}{\scriptstyle \rm}
\newcommand{\emax}{\slashed E_{\mysmall max}}
\renewcommand{\Im}{\mathrm{Im}}

\begin{document}

\title{\boldmath Next-to-leading order prediction for the decay $\mu \to e \, (e^+e^-) \, \nu \bar{\nu}$}
\author[a]{M.~Fael}\emailAdd{fael@itp.unibe.ch}
\author[a]{C.~Greub} \emailAdd{greub@itp.unibe.ch}
\affiliation[a]{Albert Einstein Center for Fundamental Physics, Institute
  for Theoretical Physics,\\ University of Bern, CH-3012 Bern,
  Switzerland.}

\date{\today}

\abstract{
We present the differential decay rates and the branching ratios of the muon decay with internal conversion, $\mu \to e \, (e^+e^-) \, \nu \bar{\nu}$, in the Standard Model at next-to-leading order (NLO) in the on-shell scheme.
This rare decay mode of the muon is among the main sources of background to the search for  $\mu \to eee$ decay.
We found that in the phase space region where the neutrino energies are small, and the three-electron momenta have a similar signature as in the $\mu \to eee$ decay, the NLO corrections decrease the leading-order prediction by about $10-20\%$ depending on the applied cut.}

\maketitle
\flushbottom

\section{Introduction}

Lepton flavour is not a conserved quantity in nature, indeed neutrino oscillations have unveiled that the Standard Model (SM) must be modified to include neutrino masses and mixing.
Processes with charged lepton flavour violation (CLFV) can occur, therefore, through neutrino mixing in the loops; however, their rates are extremely small because of the suppression given by neutrino masses. For example the $\mu \to e \gamma$ branching ratio, which is proportional to $(m_\nu/M_W)^4$, is estimated to be at the level of $10^{-50}$ or smaller, far beyond the sensitivity of any foreseeable experiment.
For this reason, an experimental observation of CLFV would be a bright evidence of new physics (NP) beyond the SM.

Muon and tau decays with CLFV have been studied in a model independent way in the framework of higher dimensional operators~\cite{Dassinger:2007ru,Crivellin:2013hpa,Pruna:2015jhf} and in a general effective field theory description of the weak interactions at low energies~\cite{Flores-Tlalpa:2015vga}.
Many scenarios of NP introduce additional sources of mixing between the lepton families that can easily lead to strong CLFV contributions.
CFLV violation can be realized, for example, in minimal see-saw type extensions of the SM~\cite{Ilakovac:1994kj,Dinh:2012bp}, in the MSSM via flavour non-diagonal SUSY breaking terms~\cite{Borzumati:1986qx,Brignole:2004ah,Paradisi:2005fk,Altmannshofer:2009ne,Girrbach:2009uy}, or in the context of two-Higgs-doublet models~\cite{Paradisi:2005tk,Crivellin:2013wna,Crivellin:2015hha,Omura:2015xcg} and composite Higgs models~\cite{Feruglio:2015gka} with generic flavour structures in the lepton sector.

Despite the tau lepton has the advantage of a mass greater than the muon one and thus, from the theoretical point of view, a better sensitivity to NP effects, muons can be produced in a much larger quantity and can be measured with much better sensitivity.
In the muon sector, the upcoming~\textsc{Mu2e}~\cite{Bartoszek:2014mya}, \textsc{DeeMe}~\cite{Natori:2014yba} and \textsc{Comet}~\cite{Kuno:2013mha} experiments will search for $\mu \to e$ conversion of muons bounded to a nucleus, while \textsc{Meg}~\cite{TheMEG:2016wtm} and \textsc{Mu3e}~\cite{Blondel:2013ia} at PSI are dedicated to the SM forbidden decays $\mu \to e \gamma$ and $\mu \to eee$, respectively.

The current limits on the $\mu \to e$ transitions are very stringent due to the constraints from \textsc{Meg} and \textsc{Sindrum} collaborations:
\begin{align}
  \mathcal{B}(\mu^+ \to e^+\gamma) &< 4.2 \times 10^{-13} \quad 90\% \text{ C.L. \cite{TheMEG:2016wtm}},\\
  \mathcal{B}(\mu^- \to e^- e^+ e^-) &< 1.0 \times 10^{-12} \quad 90\% \text{ C.L. \cite{Bellgardt:1987du}}.
\end{align}
These bounds will be further improved in the future: \textsc{Meg-II} is planned to measure the $\mu \to e \gamma$ branching ratio down to $4\times 10^{-14}$~\cite{Baldini:2013ke} while \textsc{Mu3e} is expected to reach the level of $10^{-16}$ for $\mu \to eee$~\cite{Blondel:2013ia,Berger:2014vba,Perrevoort:2016nuv}.

Such an unprecedented precision requires an equivalent level in the control and suppression of the background.
Radiative muon decay, $\mu \to e \gamma \nu \bar{\nu}$, constitutes an important source of background to $\mu \to e \gamma$ searches; secondly it provides a tool for calibration, normalization and quality check of the experiment~\cite{Adam:2013gfn}.
It was studied at next-to-leading order (NLO) accuracy in~\cite{Fael:2015gua,Fael:2016hnz}.
The background in $\mu \to eee$ searches originates from the accidental coincidence of normal muon-decay electrons and positrons, that within the detector resolution show the characteristics of the decay signal, and from the muon decay with internal conversion,  
$\mu \to e \, (e^+ e^-) \, \nu \bar{\nu}$,
which is indistinguishable from the signal except for the energy carried out by neutrinos.
This SM background can be suppressed only via an excellent momentum resolution  and a precise reconstruction of the three-electron total energy, which must be as close as possible to the muon mass.

While the SM decay rate of 
$\mu \to e \, (e^+ e^-) \, \nu \bar{\nu}$
was studied at the leading order (LO) in \cite{Bardin:1972qq,Fishbane:1985xz,vanRitbergen:1999fi,Djilkibaev:2008jy}, radiative corrections are currently missing in the literature.
The precision goal of \textsc{Mu3e} experiment and the expected momentum resolution, about $0.5$ MeV~\cite{Berger:2014vba,Perrevoort:2016nuv}, call for the estimate of the NLO corrections, especially in the narrow region of the phase space where the missing energy is small and the three-electron momenta have a similar signature as the $\mu \to eee$ decay.
It is the aim of this paper to present the first calculation of such NLO corrections.\footnote{In the last phase of this work we learnt about an ongoing independent calculation of the NLO corrections to $\mueee$ decay carried out by G.\ M.\ Pruna, A.\ Signer and Y.\ Ulrich~\cite{Pruna:2016spf}.}

We begin our analysis in section~\ref{sec:lo} reviewing the SM prediction for the differential decay rate at LO.
The technical ingredients employed in our calculation of virtual and real corrections are presented in section~\ref{sec:nlo}.
Our NLO predictions for the branching ratios are reported in section~\ref{sec:br}, where we discuss also the impact on CLFV searches. Conclusions are drawn in section~\ref{sec:con}.

\section{Conventions and LO decay rate}
\label{sec:lo}
\begin{figure}[h]
  \centering
  \includegraphics[width=0.8\textwidth]{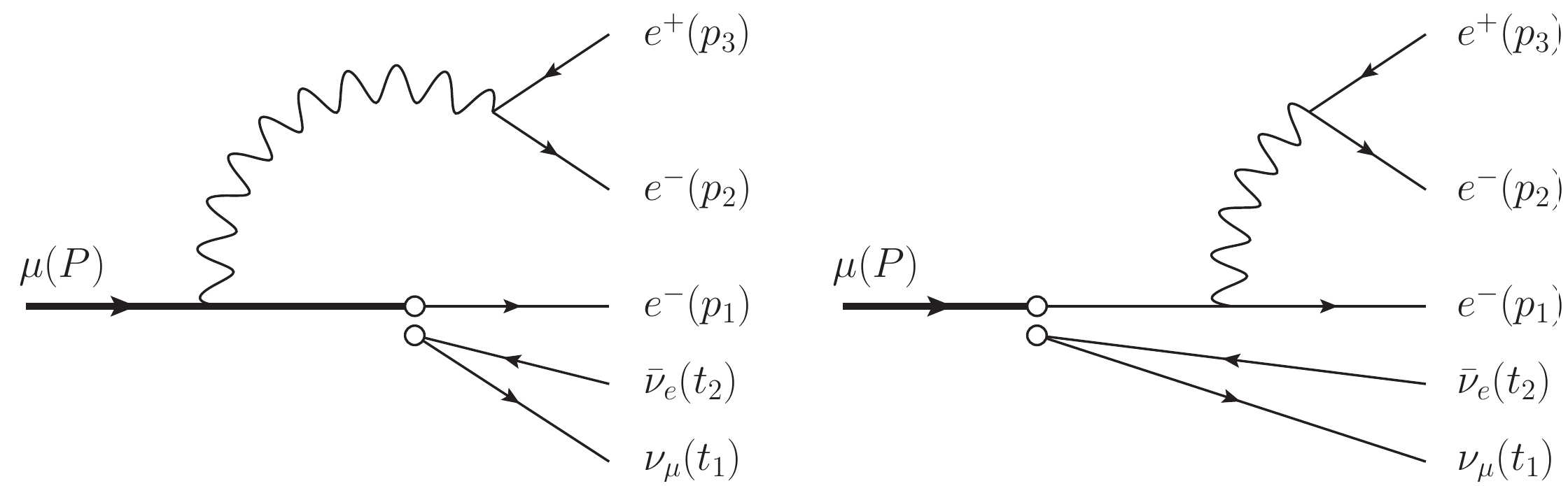}
  \caption{Feynman diagrams for $\mueee$ in the Fermi $V$--$A$ effective theory. Two other diagrams with $p_1$ and $p_2$ interchanged must be considered also.}
  \label{fig:LO}
\end{figure}

In this section we introduce our conventions and we discuss the tree-level decay rate.
In the SM the decay of a muon into three electrons and two neutrinos proceeds through the emission of an off-shell photon with subsequent internal conversion into $e^+e^-$, as shown in figure~\ref{fig:LO}.
Since we are interested only in the leading contribution in $G_F$, photon radiation from the $W$ boson and electron-pair production from heavy bosons will not be considered in the following.

Let us consider specifically the decay of a negative muon:
\begin{equation}
  \mueee
  \label{eqn:mueee}.
\end{equation}
In the SM, the tree-level decay rate for an unpolarized muon is, in its rest frame,
\begin{equation}
  \frac{d^6 \Gamma}
  {d t  \,d m_{123} \, d m_{12} \, 
  d\! \cos \theta_{3}^{*} \, d \Omega_{1}^{**} } = 
  \frac{\alpha^2 G_F^2}{96 \pi^6 m^2_\mu} \, 
  G_{\mysmall LO}(t,m_{123},m_{12},\cos\theta_3^{*},\Omega_1^{**}) \,
  t \,
  \vert \vec{p}_{123}\vert
  \vert \vec{p}_3^{\,*}\vert\,
  \vert \vec{p}_1^{\, **} \vert,
  \label{eqn:LOdecayrate}
\end{equation}
where 
$G_F=1.166 \, 378 \, 7(6) \times10^{-5}$ GeV$^{-2}$~\cite{Webber:2010zf} 
is the Fermi constant, defined from the muon lifetime, and
$\alpha = 1/137.035\,999\,157\,(33)$
is the fine-structure constant~\cite{Aoyama:2012wj,Aoyama:2014sxa}.
Calling $m_\mu$ and $m_e$ the masses of the muon and the electron (neutrinos are considered to be massless), we define the ratio $r=m_e/m_\mu$.
The four-vectors $P$, $p_1$, $p_2$ and $p_3$ are the momenta of the muon, the two electrons and the positron, respectively. 
We denote with $p_{123} = p_1+p_2+p_3$ the sum of the electrons and the positron momenta and with $m_{123}^2=p_{123}^2$ their invariant mass squared.
Also, we define $m_{12}^2=(p_1+p_2)^2$ and $t^2=(t_1+t_2)^2=(P-p_{123})^2$ the invariant masses squared of the two electrons and the two neutrinos, respectively --- $t_1$ and $t_2$ are the neutrino and anti-neutrino momenta.
Note that the dependence on the (undetected) neutrino momenta has been integrated out analytically in~\eqref{eqn:LOdecayrate}. 

Solid angles and momenta labeled with the superscript `$*$' are in the center-of-mass system (c.m.s.) of the two electrons and the positron, where $p_{123}^* = (m_{123},\vec{0})$, while those with `$**$' are in the c.m.s.\ of the two electrons, where $p_{12}^{**} = (m_{12},\vec{0})$.
The $z$-direction of the solid angles $d\Omega_3^*$ and $d\Omega_1^{**}$ are given by the direction of $\vec{p}_{123}$ and $\vec{p}_{12}=\vec{p}_1+\vec{p}_2$, respectively.

The dimensionless quantity $G_{\mysmall LO}$ is a rational function proportional to the Born squared matrix element --- it is thus Lorenz invariant --- and it depends on all possible scalar products built with the momenta $P,p_1,p_2$ and $p_3$.  We introduced such a function in order to factorize out the physical constants $\alpha$, $G_F$ and $m_\mu$ from the squared matrix element. 
Two indistinguishable $e^-$ are present in the final state; the symmetry property of the matrix element assures that $G_{\mysmall LO}$ and the decay rate~\eqref{eqn:LOdecayrate} are symmetric under the exchange $p_1 \leftrightarrow p_2$.

The (lengthy) explicit expression of eq.~\eqref{eqn:LOdecayrate} is provided as an ancillary file of this paper. For further details, we refer the reader to our Appendix.

\section{NLO corrections: details of the calculation}
\label{sec:nlo}
In this section we will consider the SM prediction for the differential decay rate of~\eqref{eqn:mueee} at NLO in $\alpha$.
Virtual and real corrections are evaluated in the Fermi $V$--$A$ effective theory of weak interactions:
\begin{equation}
  \mathcal{L} = 
  \mathcal{L}_\text{QED} 
  +\mathcal{L}_\text{QCD} 
  +\mathcal{L}_\text{Fermi}.
  \label{eqn:lfermi}
\end{equation}
The Fermi Lagrangian is
\begin{equation}
  \mathcal{L}_{\mysmall Fermi} =
  -\frac{4G_F}{\sqrt{2}}
  (\bar{\psi}_{\nu_\mu} \gamma^\alpha P_L \psi_\mu) \cdot
  (\bar{\psi}_e \gamma_\alpha P_L \psi_{\nu_e}) + \text{h.c.} \, ,
  \label{eqn:LFermi}
\end{equation}
where $\psi_\mu, \psi_e,\psi_{\nu_\mu},\psi_{\nu_e}$ denote the fields of the muon, the electron and their associated neutrinos, respectively; $P_L =(1-\gamma_5)/2$ denotes the left-hand projector operator.
Under this approximation tiny term of $\mathcal{O}(\alpha m_\mu^2/M_W^2) \sim 10^{-8}$ due to the finite $W$-boson mass are neglected ---  they are even smaller than the NNLO corrections of $\mathcal{O}(\alpha^2)$.

A Fierz rearrangement of the four-fermion interaction~\eqref{eqn:LFermi} allows us to factorize the amplitudes of virtual and real corrections into the product of spinor chains depending either on the neutrino momenta or on the muon and electron ones.
In this way the neutrino phase space integral can be carried out analytically, as was done at tree level, so as to decrease by two units the dimensionality of the integral that must be performed numerically.

\subsection{Virtual corrections}
\begin{figure*}[ht]
  \centering
  \includegraphics[width=0.45\textwidth,valign=t]{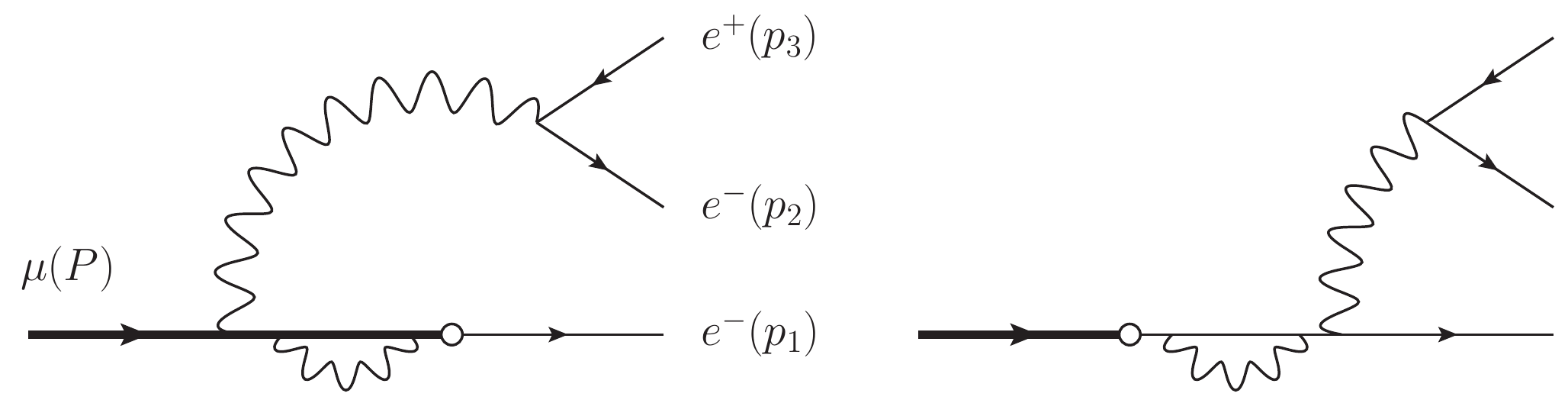}\qquad
  \includegraphics[width=0.45\textwidth,valign=t]{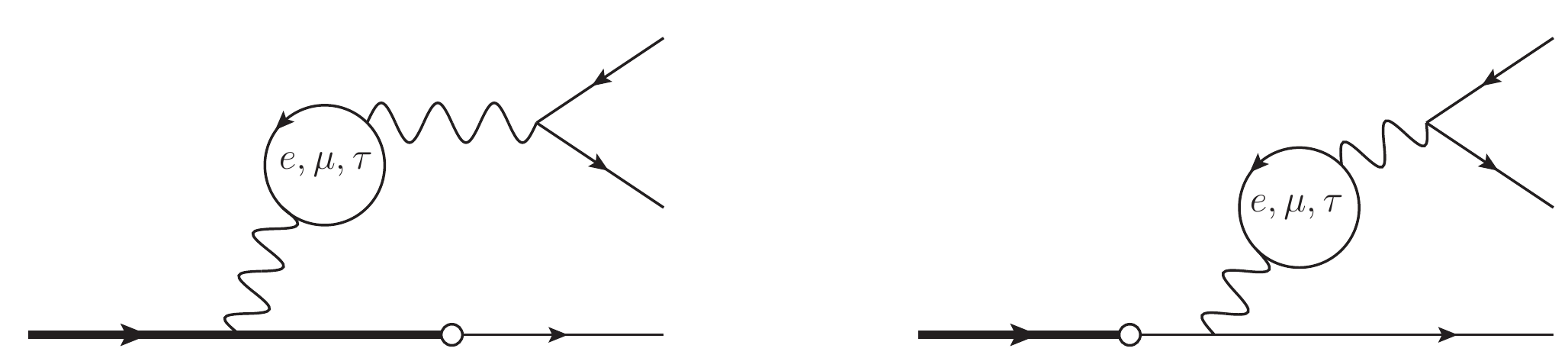}\\
  \includegraphics[width=0.45\textwidth,valign=t]{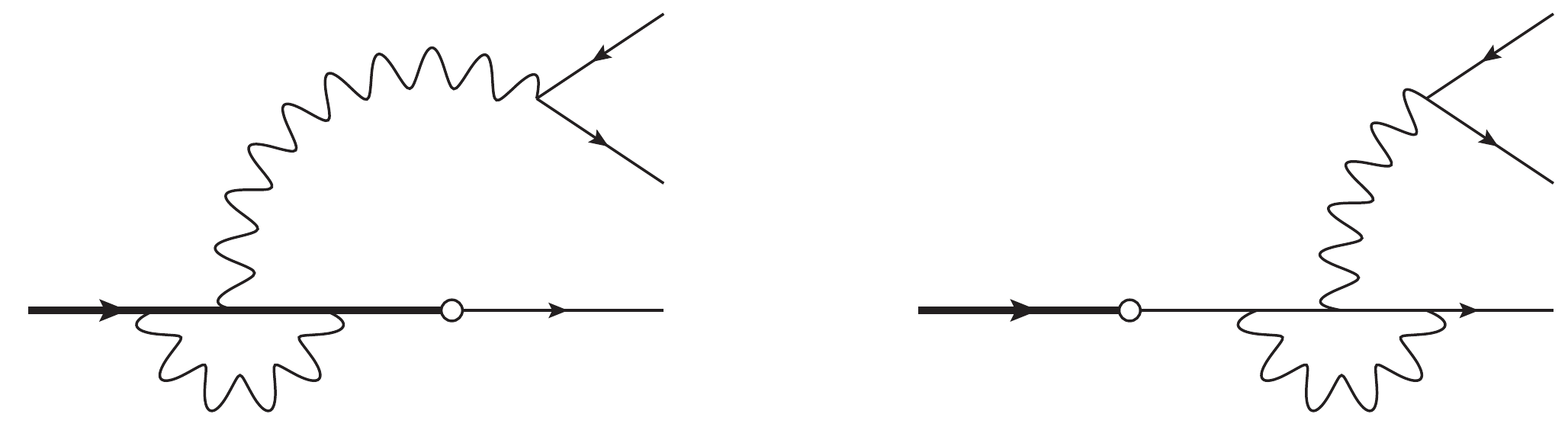}\qquad
  \includegraphics[width=0.45\textwidth,valign=t]{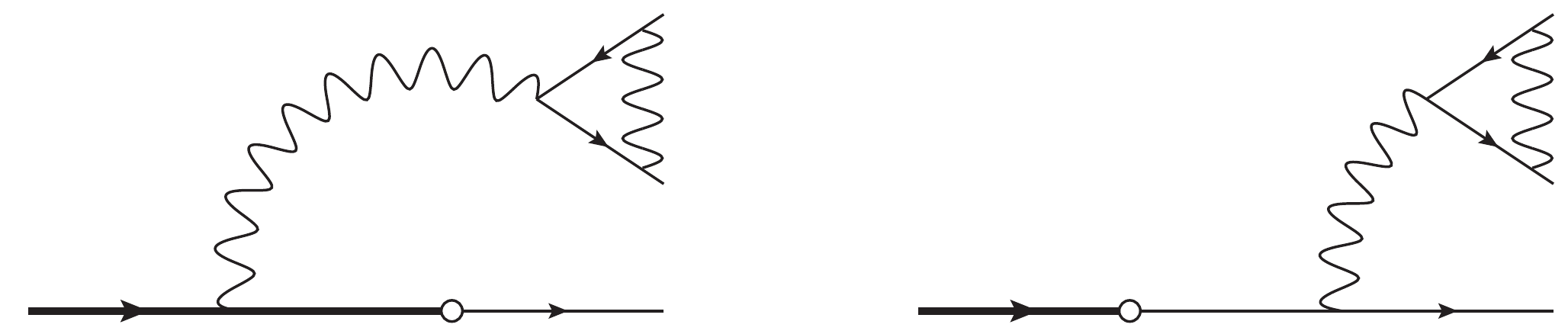}\\
  \includegraphics[width=0.45\textwidth,valign=t]{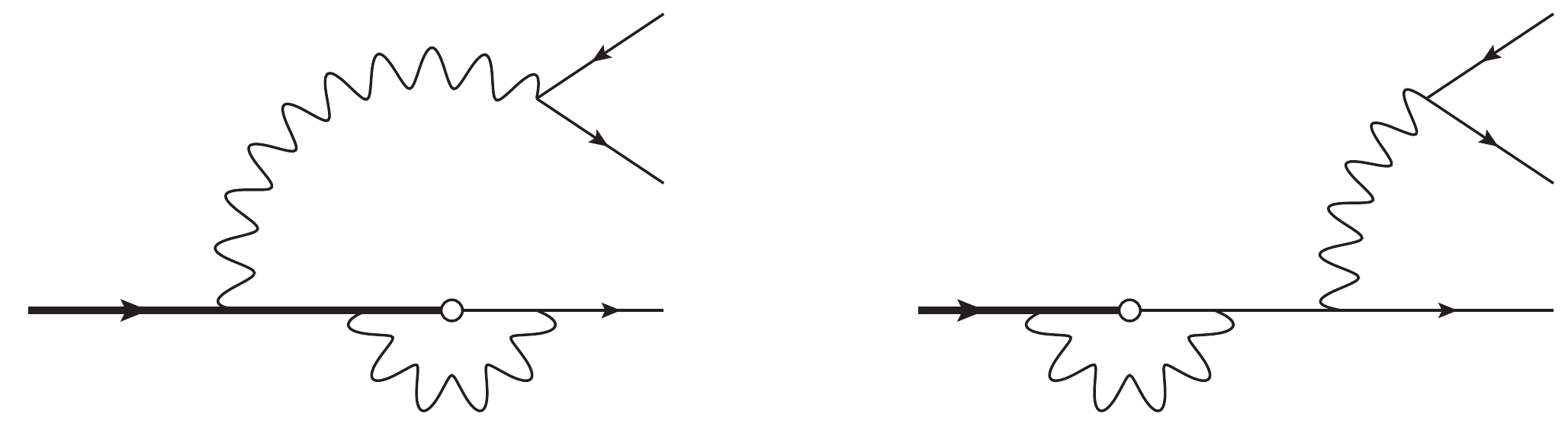}\qquad
  \includegraphics[width=0.45\textwidth,valign=t]{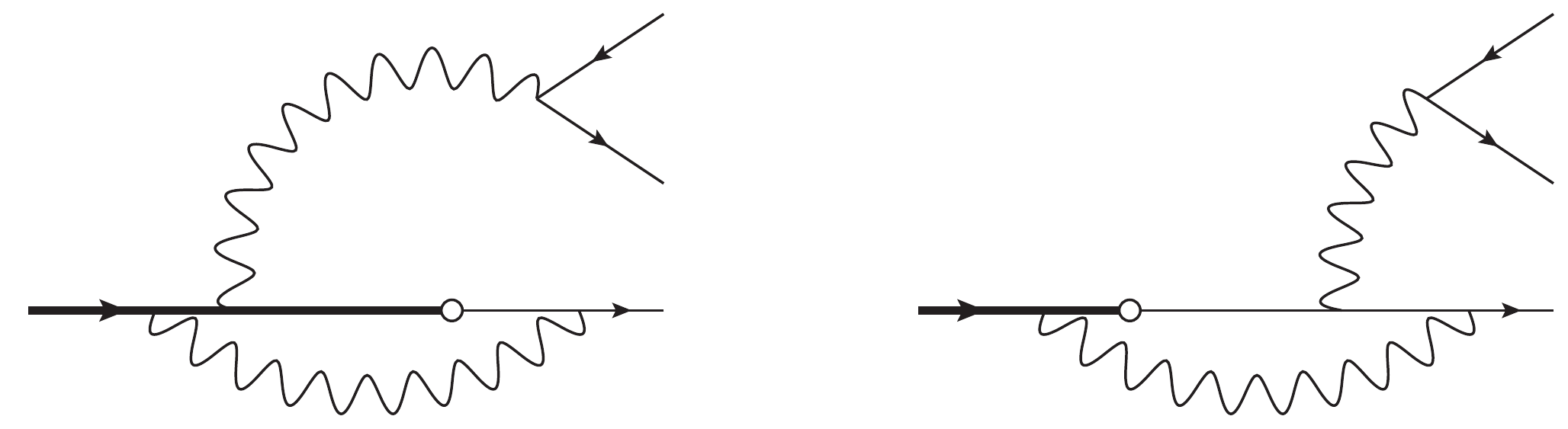}\\
  \includegraphics[width=0.45\textwidth,valign=t]{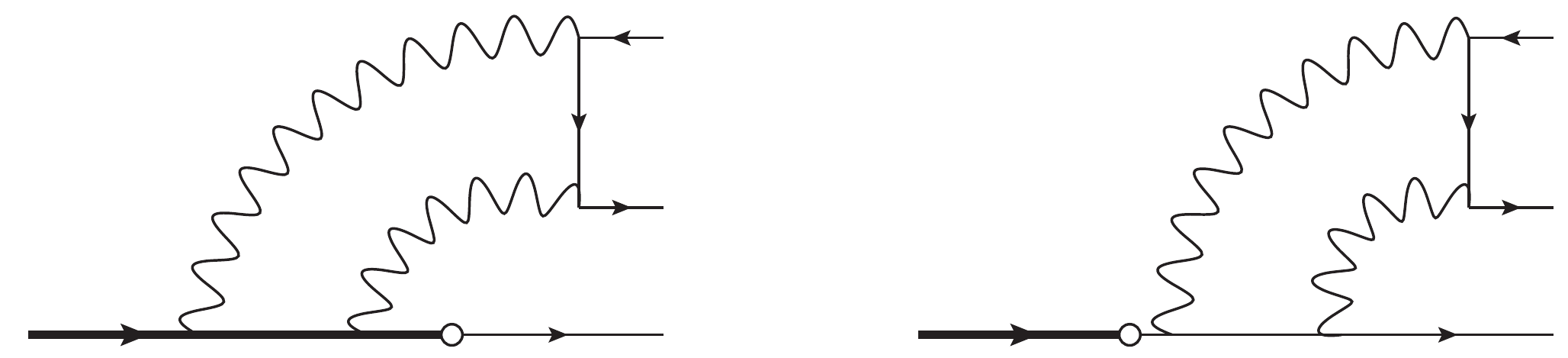}\qquad
  \includegraphics[width=0.45\textwidth,valign=t]{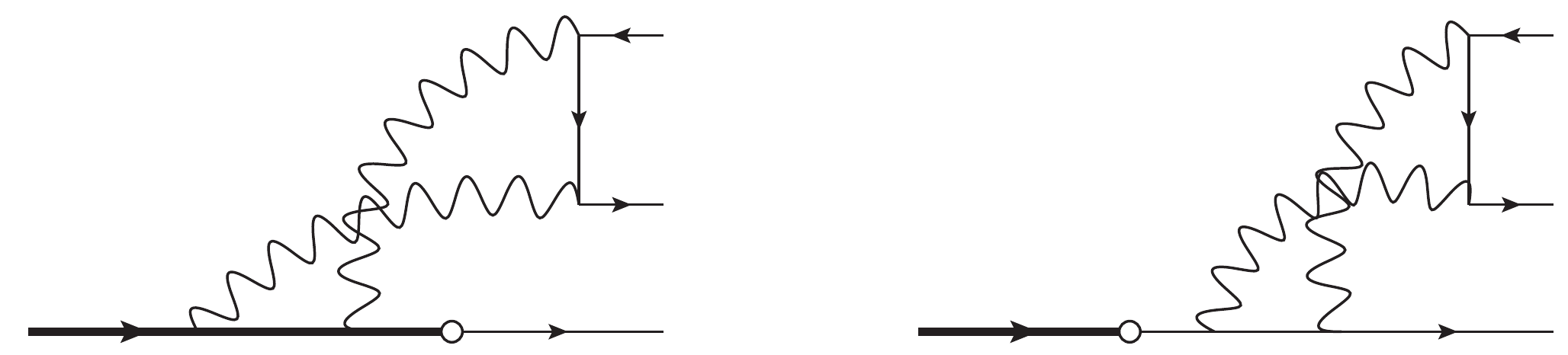}\\
  \includegraphics[width=0.45\textwidth,valign=b]{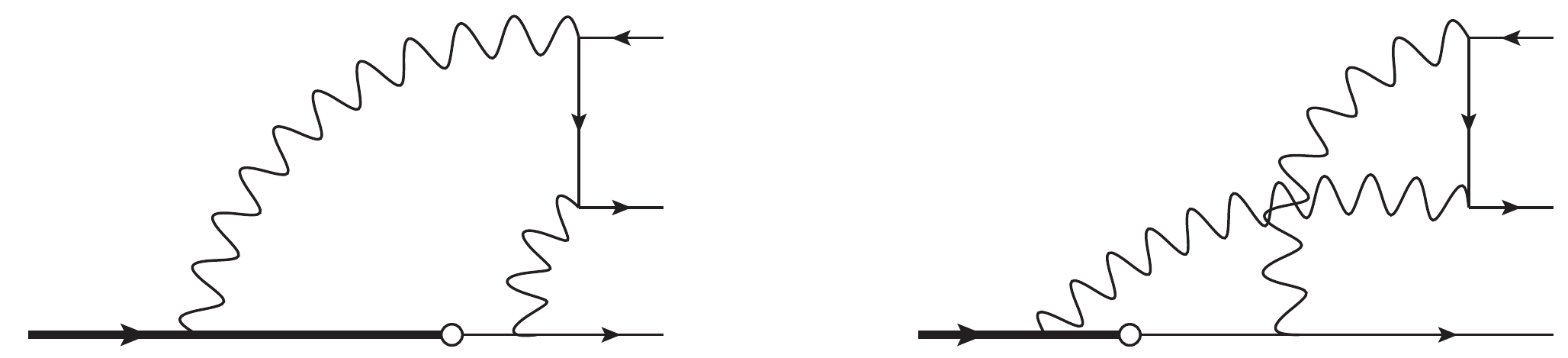}\qquad
  \includegraphics[width=0.45\textwidth,valign=b]{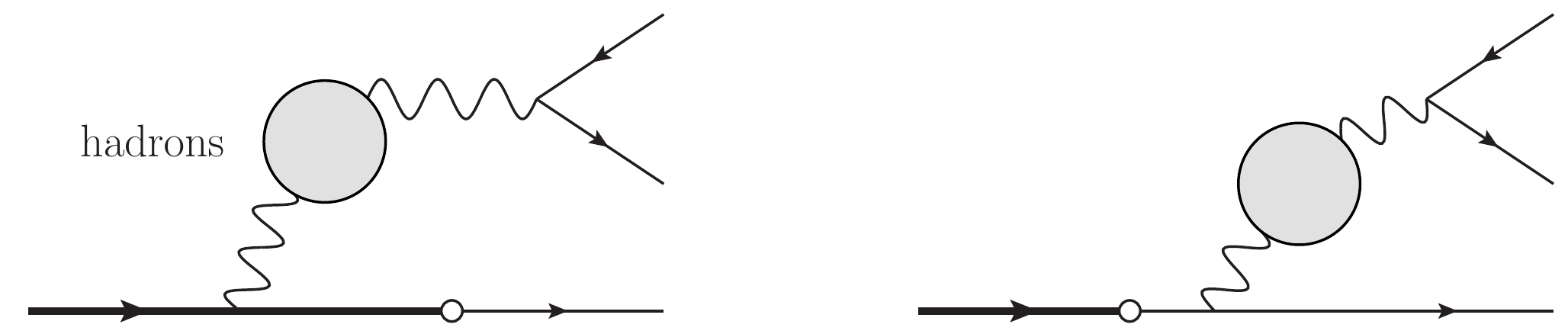}\\
  \caption{One-loop diagrams for $\mueee$ decay. 
  The muon and the electrons are drawn with bold and thin lines, respectively; the dots symbolize the Fermi interaction (for simplicity the neutrinos are not drawn). For each diagram, a symmetric one with $p_1 \leftrightarrow p_2$ interchanged must be considered. }
  \label{fig:oneloop}
\end{figure*}

The one-loop amplitudes, shown in figure~\ref{fig:oneloop}, are reduced to tensor integrals and subsequently decomposed into their Lorentz-covariant structure by means of the algebra manipulation program \textsc{Form}~\cite{Kuipers:2012rf} and the \emph{Mathematica} package \textsc{FeynCalc}~\cite{Mertig:1990an,Shtabovenko:2016sxi}.
For the numerical evaluation of the tensor-coefficient functions we employed the \textsc{LooopTools} library~\cite{Hahn:1998yk,vanOldenborgh:1989wn}.

Ultraviolet (UV) divergences are regularized via dimensional regularization; UV-finite results are obtained by renormalizing the theory~\eqref{eqn:lfermi} in the on-shell scheme. Indeed, as shown long ago by Berman and Sirlin~\cite{BermanSirlin1962}, to leading order in $G_F$, but to all orders in $\alpha$, the radiative corrections to muon decay in the Fermi $V$--$A$ theory are finite after fermion mass and wave function renormalization. A small photon mass $\lambda$ is introduced to regularize the infrared (IR) divergences, while the finite electron mass $m_e$ regularizes the collinear ones. 

The contribution to the rate coming from the hadronic vacuum polarization, which is not calculable at low energy in perturbative QCD, is quite small in the muon case since the invariant mass of the electron-positron pair never exceeds the pion threshold. However this kind of correction starts to be relevant if one considers, instead of the muon, the tau decays $\tau \to \ell (\ell^+\ell^-) \nu \bar{\nu}$.
These effects can be taken into account expressing the hadronic vaccum polarization, $\Pi_{\mysmall had}(q^2)$, in terms of $e^+ e^- \to \text{hadrons}$ cross section data: 
\begin{equation}
  R_{\mysmall had}(s) = \sigma(e^+ e^- \to \, \text{hadrons})/
  \frac{4 \pi \alpha(s)^2}{3s}.
\end{equation}
The normalization factor $4 \pi \alpha(s)^2/(3s)$ is the tree-level cross section of $e^+e^- \to \mu^+ \mu^-$ in the limit $s \gg 4m_\mu^2$ --- note that $\sigma(e^+ e^- \to \, \text{hadrons})$ does not include initial state radiation or vacuum polarization corrections.
The optical theorem connects $R_{\mysmall had}(s)$ to the imaginary part of hadronic vacuum polarization:
\begin{equation}
  \Im \, \Pi_{\mysmall had} (s) =
  \frac{\alpha(s)}{3} 
  R_{\mysmall had} (s).
\end{equation}
The vacuum polarization can be then obtained by means of the dispersion relation.
In this work, we made use of the package \textsc{alphaQED}~\cite{Fred,Jegerlehner:2001ca,Jegerlehner:2006ju,Jegerlehner2011a} for the evaluation of the functions $R_{\mysmall had}$ and $\Pi_{\mysmall had} $.

\subsection{Real photon emission}
The rate of the bremsstrahlung process, the decay \eqref{eqn:mueee} where an additional photon is produced,  blows up when the photon energy becomes small, leading in the phase space integral to the well-known logarithmic IR singularity. 
In order to handle the IR singularity we adopted a phase-space slicing method: we introduced a small photon energy cut-off $\omega_0$ and we divided the real emission contribution into a \emph{soft} and a \emph{hard} part.

The \emph{soft} part, which contains the IR singularity, comes from the phase space region where the photon energy is below the threshold $\omega_0$. As in the case of the virtual diagrams, the IR singularity is regularized by a small photon mass $\lambda$.
By taking advantage of the soft photon approximation for the amplitude of the bremsstrahlung process, it is possible to perform the integral with respect to the photon momentum analytically; the process-independent result, which was derived in~\cite{'tHooft:1978xw} (see also ref.~\cite{Denner:1991kt}), depends only on the charges and momenta of external particles in the corresponding Born process. 
We checked that the IR divergence cancels out once the soft part is added to the one-loop diagrams. 

Although the addition of the soft photon emission to the virtual corrections is sufficient to obtain a finite differential width, in general it is not adequate for real experiments since they cannot provide a photon energy threshold $\omega_0$ small enough for the validity of the soft photon approximation, which neglects terms of order $\omega_0/m_\mu$.
Therefore it is necessary to include the \emph{hard} part as well, i.e.\ the contribution to the rate due to photons with energy greater than $\omega_0$. 
The soft and the hard parts must be properly merged to assure the $\omega_0$-independence of the final physical observables: 
the numerical integration of hard part yields a $\log \omega_0/m_\mu$ enhancement that must cancel against the explicit $\omega_0$-dependence in the soft part for sufficiently small values of $\omega_0$.
The value of $\omega_0 $ can be fixed once the prediction for the NLO corrections reaches a sort of ``plateau'', i.e.\ when a further reduction of $\omega_0 $ cannot be resolved anymore within the numerical error.
\section{Results}
\label{sec:br}
\subsection{Branching fraction}
The branching ratio $\mathcal{B}$ of the $\mueee$ decay can be obtained integrating the differential decay rate~\eqref{eqn:LOdecayrate} over the allowed kinematic ranges,
\begin{align}
  -1 & \le \cos \theta_{3}^{*} \le 1, & 
  3 m_e & \le m_{123} \le m_\mu, \notag\\
  -1 & \le \cos \theta_{1}^{**} \le 1, &
  0 & \le t \le m_\mu -m_{123}, \notag \\
  0 & \le \varphi_1^{**} < 2\pi, &
  2 m_e &\le m_{12} \le m_{123}-m_e,
  \label{eqn:integrationlimits}
\end{align}
and multiplying it by the muon lifetime $\tau_\mu = 2.1969811 \, (22) \times 10^{-6}$ s~\cite{Olive:2016xmw}.
We recall that the Particle Data Group (PDG) defines the Fermi constants of weak interactions from the muon lifetime evaluated in the Fermi $V$--$A$ theory~\cite{Olive:2016xmw}; its definition is given by
\begin{equation}
  \frac{\hbar}{\tau_\mu} = 
  \frac{G_F^2 m_\mu^5}{192 \pi^3}
  F(r^2)
  \left( 
    1+\delta_\mu  \right),
  \label{eqn:tmuon}
\end{equation}
where
$
  F(x) = 1-8x+8x^3-x^4-12x^2 \ln x
$
is the phase space factor while $\delta_\mu$ incorporates the QED correction evaluated in the Fermi $V$--$A$ theory: the corrections of virtual and real photons up to ${\cal O} (\alpha^2)$, as well as the contribution of the decay \eqref{eqn:mueee} at tree level~\cite{Behrends:1955mb,Berman:1958ti,Kinoshita:1958ru,Roos:1971mj,vanRitbergen:1998yd,Steinhauser:1999bx,Pak:2008qt,vanRitbergen:1998hn,Ferroglia:1999tg}.
Note that it is possible to make use of eq.~\eqref{eqn:tmuon}, instead of the experimental value of $\tau_\mu$, for the normalization of width; in such way the dependence on $G_F$ and $m_\mu^5$ is removed from the branching ratios. 

The analytic integration over the kinematic ranges~\eqref{eqn:integrationlimits} of the LO differential rate in eq.~\eqref{eqn:LOdecayrate} yields~\cite{vanRitbergen:1999fi}
\begin{align}
  \mathcal{B}_{\mysmall LO} &=
  \frac{\alpha^2/\pi^2}{F(r^2)(1+\delta_\mu)} \Bigg\{
    -\frac{25361}{5184}
    +\frac{25}{9} \zeta(2) 
    +\ln 2 \left( \frac{37}{216}-\frac{2}{3}\zeta(2) \right)
    -\frac{2}{9}\left( 1+ \ln 2 \right) \ln^2 2 \notag \\
    &+\frac{11}{3} \zeta(3) -\left( \frac{25}{24}-\zeta(2) \right) \ln r^2
    +I(r^2) \Bigg\},
    \label{eqn:brloanalytic}
\end{align}
where
\begin{equation}
  I(x)   = \frac{i}{2} \frac{1-x}{\sqrt{x}} 
  \int_0^1
  \left[ 2+x+(1-x) v^2 \right]
  \sqrt{1-v^2}
  K \left( iv\sqrt{\frac{1-x}{x}} \right) dv ,
  \label{eqn:kernelK}
\end{equation}
with $I(r^2) = 9.47056$, and the kernel function $K(u)$ is given in eq.~(11) of ref.~\cite{vanRitbergen:1998hn}. 
As discussed in~\cite{vanRitbergen:1999fi}, the integral~\eqref{eqn:kernelK} behaves like $\ln^n r$ for non-negative integers $n \le 3$, so that it is singular in the limit $m_e\to0$. 
It contains also vanishing terms in that limit, but since the original calculation of $K(u)$  neglected them, the result in eq.~\eqref{eqn:brloanalytic} is correct only in the terms that do not vanish as $m_e \to 0$. With the analytic result~\eqref{eqn:brloanalytic} we obtain the following prediction for LO branching ratio: $\mathcal{B}_{\mysmall LO} = 3.40 \times 10^{-5}$. This value is in agreement with the result of our numerical integration in table~\ref{tab:br}; the difference between these two values, approximately $5\%$, is due to the aforementioned terms neglected in the analytic formula~\eqref{eqn:brloanalytic}.

We present in the first row of table~\ref{tab:br} the LO and NLO branching ratios, the $O(\alpha)$ correction coming from virtual and real diagrams, denoted with $\delta \mathcal{B}_\text{NLO}$, and the $K$-factor, which is the ratio between the NLO and the LO prediction.
They are computed taking into account the full dependence on the mass ratio $r$.
The numerical integrations were performed with Monte Carlo methods by means of the \textsc{Cuba} library~\cite{Hahn:2004fe}; the results were tested with different numerical integration methods.
In table~\ref{tab:br} the uncertainty due to numerical errors is labeled with the subscript ``$n$''. Moreover we also estimated the theoretical uncertainty associated to the renormalization scale variation; they are denoted with the subscript ``$\mu$''. It is quantified by converting the renormalization scheme for $\alpha$ from the on-shell scheme to the hybrid $\overline{\rm MS}$ adopted in the NNLO calculation of the muon lifetime~\cite{vanRitbergen:1998hn,vanRitbergen:1998yd}. In this scheme the electron loop in the photon vacuum polarization is renormalized in $\overline{\rm MS}$ while all other fermion loops in the on-shell scheme. The quoted uncertainty is the difference between the NLO prediction evaluated at the renormalization scales $\mu = m_e$, corresponding to the on-shell scheme, and $\mu = m_\mu$. 

It is interesting to note that a Born-virtual interference term yields a null contribution after phase space integration.
Indeed, let us consider the one-loop diagrams where the positron-electron pair is produced by two virtual photons, corresponding to the boxes and the pentagons in the last two rows of figure~\ref{fig:oneloop}.  
If we regard  as a cut diagram the intereference between these one-loop amplitudes and the tree-level ones in figure~\ref{fig:LO}, we recognize that there is a closed fermion loop attached to three photon lines. The Furry theorem assures that the contribution after phase space integration is zero. 
However, such interference term cannot be neglected in the differential decay rate, in fact the cancellation happens between couples of phase space points related by an exchange of the $p_1$ and $p_3$ momenta (or $p_2$ and $p_3$). We explicitly verified that this interference term vanishes within the numerical error but we neglected it in our Monte Carlo integration in order to speed up the numerical convergence.

The branching ratio of~\eqref{eqn:mueee} was measured long ago by the \textsc{Sindrum} experiment~\cite{Bertl:1985mw},
\begin{equation}
  \mathcal{B}_{\mysmall EXP} (\mu^- \to e^- e^+ e^- \nu_\mu \bar{\nu}_e) = 3.4 \, (4) \times 10^{-5}.
\end{equation}
This measurement agrees with our theoretical prediction in table~\ref{tab:br}; new more precise results are expected in the future by the \textsc{Mu3e} experiment~\cite{Blondel:2013ia}.
\begin{table}[ht]
\def\arraystretch{1.2}
  \centering
  \begin{tabular}{lllll}
    $\emax$ & $\mathcal{B}_{\mysmall LO}$ & 
    $\delta \mathcal{B}_{\mysmall NLO}$ & 
    $\mathcal{B}_{\mysmall NLO}$ &
    $K$ \\ \hline
    no cut & $3.6054 \, (1)_n \times 10^{-5}$ & 
    $-6.69 \,(5)_n \times 10^{-8}$ & 
    $3.5987 \, (1)_n \, (8)_{\mu} \times 10^{-5}$ & 0.998\\
    $1\, m_e$ & 
    $2.8979 \, (6)_n \times 10^{-19}$ & 
    $ -6.56 \,(2)_n \times 10^{-20} $ & 
    $2.242 \,(2)_n \, (17)_{\mu} \times 10^{-19}$ & 
    $0.77$\\
    $5 \, m_e$ & 
    $4.641 \,(1)_n \times 10^{-15}$ &
    $-7.41 \,(3)_n \times 10^{-16}$&
    $3.900 \, (3)_n \, (20)_{\mu}  \times 10^{-15}$&
    $0.83$\\
    $10 \, m_e$ &
    $3.0704 \, (7)_n \times 10^{-13}$ &
    $-4.04 \,(2)_n  \times 10^{-14}$&
    $2.666 \,(2)_n \, (11)_{\mu} \times 10^{-13}$&
    $0.87$\\
    $20 \, m_e$ &
    $2.1186 \, (5)_n \times 10^{-11}$ &
    $-2.17 \,(1)_n  \times 10^{-12}$&
    $1.902 \,(1)_n \, (6)_{\mu} \times 10^{-11}$&
    $0.90$\\
    $50 \, m_e$ &
    $7.151 \, (1)_n \times 10^{-9}$ &
    $-4.55 \,(3)_n  \times 10^{-10}$&
    $6.696 \, (3)_n \, (13)_{\mu} \times 10^{-9}$&
    $0.93$\\
    $100\, m_e$ &
    $2.1214 \,(4)_n \times 10^{-6}$ &
    $-9.47 \, (6)_n  \times 10^{-8}$&
    $2.027 \, (1)_n \, (3)_{\mu} \times 10^{-6}$&
    $0.96$
  \end{tabular}

  \caption{LO and NLO branching ratios of $\mueee$ with and without a cut on the missing energy, the $O(\alpha)$ correction given by the sum of one-loop and real emission diagrams, $\delta \mathcal{B}_{\mysmall NLO}$, and the $K$-factor, $K=\mathcal{B}_\text{NLO}/\mathcal{B}_\text{LO}$.
The uncertainties are due to numerical error~($n$) and renormalization scale variation~($\mu$) (see the text for details).}
  \label{tab:br}
\end{table}

\subsection{Impact on CLFV searches}
\begin{figure*}[ht]
  \centering
  \subfloat[][]{\includegraphics[width=0.5\textwidth]{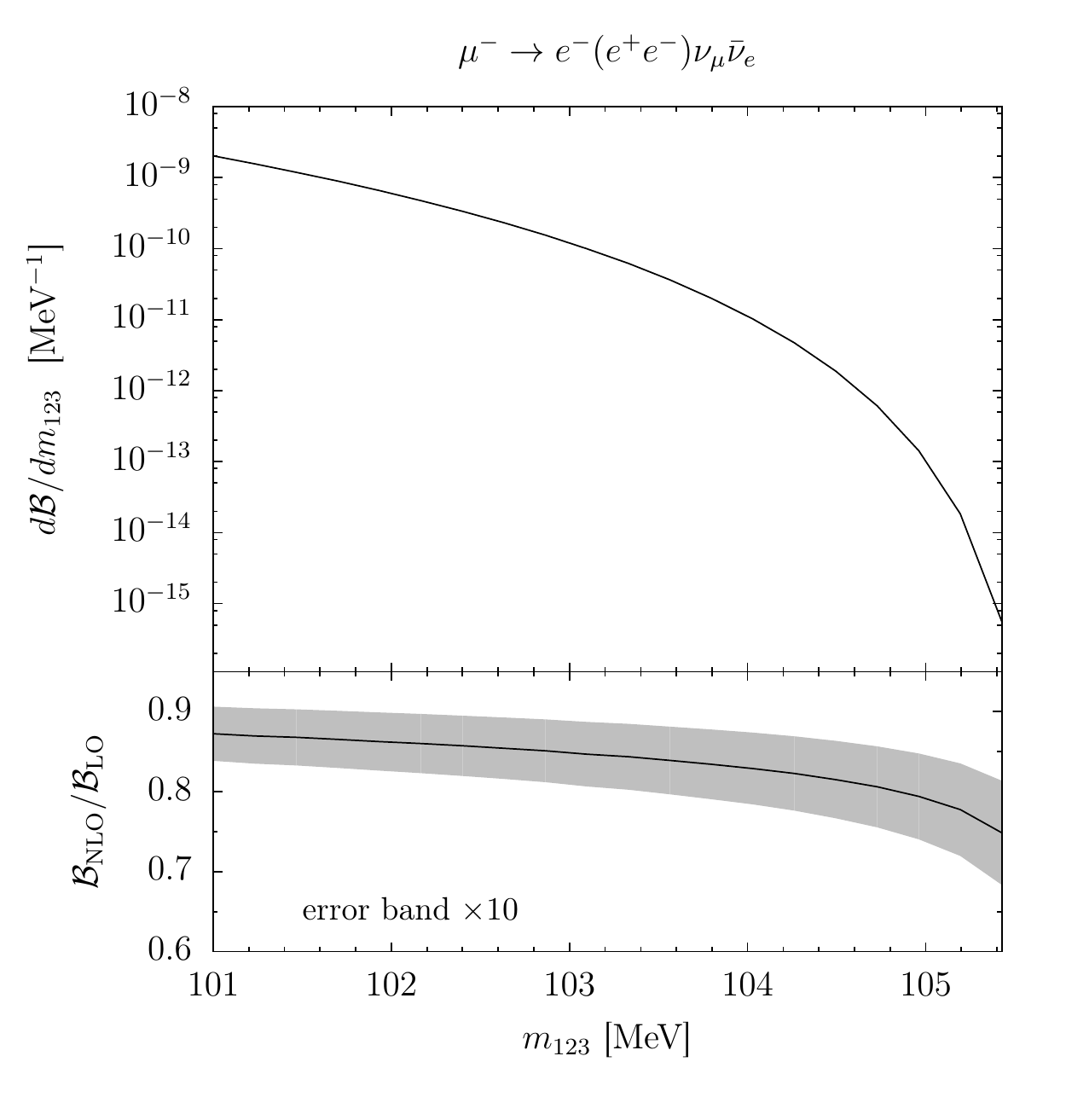}\label{fig:BRminv}}
  \subfloat[][]{ \includegraphics[width=0.5\textwidth]{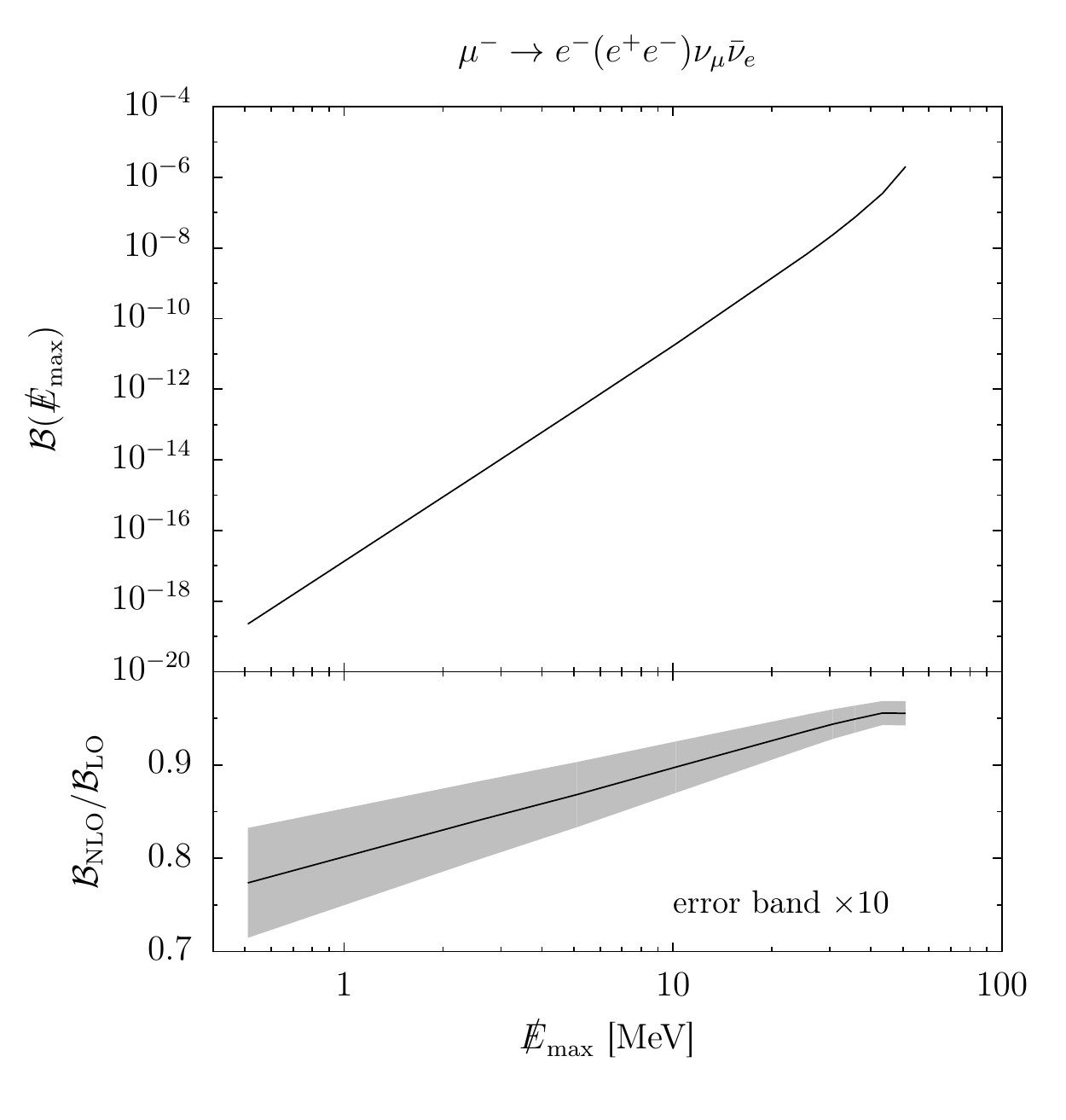}\label{fig:BRemax}
}
  \caption{The $\mueee$ branching ratio at NLO as a function of the three-electron invariant mass $m_{123}$ (left) and the invisible energy cut $\slashed E_{\rm max}$ (right).
The ratio between the NLO and LO predictions is depicted in the lower part of each panel. 
The error band (magnified 10 times) represents the assigned theoretical error due to renormalization scale variation.}
  \label{fig:brall}
  \end{figure*}

The relative magnitude of radiative corrections were also studied in the specific final-state configuration of the decays~\eqref{eqn:mueee} where the neutrino energies  are very small and the total energy of the three electrons is close to $m_\mu$. 
As already mentioned in the introduction, this phase-space region is of particular interest to $\mu \to eee $ searches because the muon decay~\eqref{eqn:mueee} can mimic the three-body decay mode with CLFV.

The upper panel of figure~\ref{fig:BRminv} shows  $d\mathcal{B}/dm_{123}$, the normalized NLO differential rate as function of the three-electron invariant mass $m_{123}$, close to the end point region $m_{123} = m_\mu$.
The local $K$-factor is drawn in the lower part. 
The rate, evaluated at fixed value of $m_{123}$, is fully inclusive in the bremsstrahlung photon.
%

Beside that, we calculated also the branching fraction applying a cut on the missing energy, in analogy to the analysis done in ref.~\cite{Djilkibaev:2008jy} at LO.
Let us define 
$\mathcal{B} (\slashed E_{\rm max})$ 
to be the integral of the differential decay rate over the phase space region satisfying 
\begin{equation}
  \slashed E = m_\mu-E_1-E_2-E_3 \le \emax.
  \label{eqn:emax}
\end{equation}
This constraint is fulfilled at LO applying, under the condition $\emax < m_\mu/2$, the following integration limits:
\begin{equation}
  0 \le t \le \emax, \qquad
  m_{123}^\text{min} \le m_{123} \le m_\mu-t,
  \label{eqn:emaxbounds}
\end{equation}
where $m_{123}^\text{min} = \sqrt{m_\mu^2-2 \emax m_\mu + t^2}$; the other limits of integration are left unchanged.
In the calculation of the NLO corrections to $\mathcal{B}(\emax)$ we assumed the maximum missing energy $\emax$ to be smaller than the photon detection threshold, and much lower than the muon mass. 
In figure~\ref{fig:BRemax} we show the branching ratio $\mathcal{B}_{\mysmall NLO}(\emax)$ versus the cut on the missing energy, in the upper panel, and its relative magnitude with respect to the LO prediction, in the lower panel.
Error bands depicted in figure~\ref{fig:brall} are the assigned theoretical error due to renormalization scale variation. They are evaluated as in the case of the inclusive branching ratio. Errors due to numerical integration are typically smaller than the first.%

\begin{figure}[htb]
  \centering
  \includegraphics[width=0.7\textwidth]{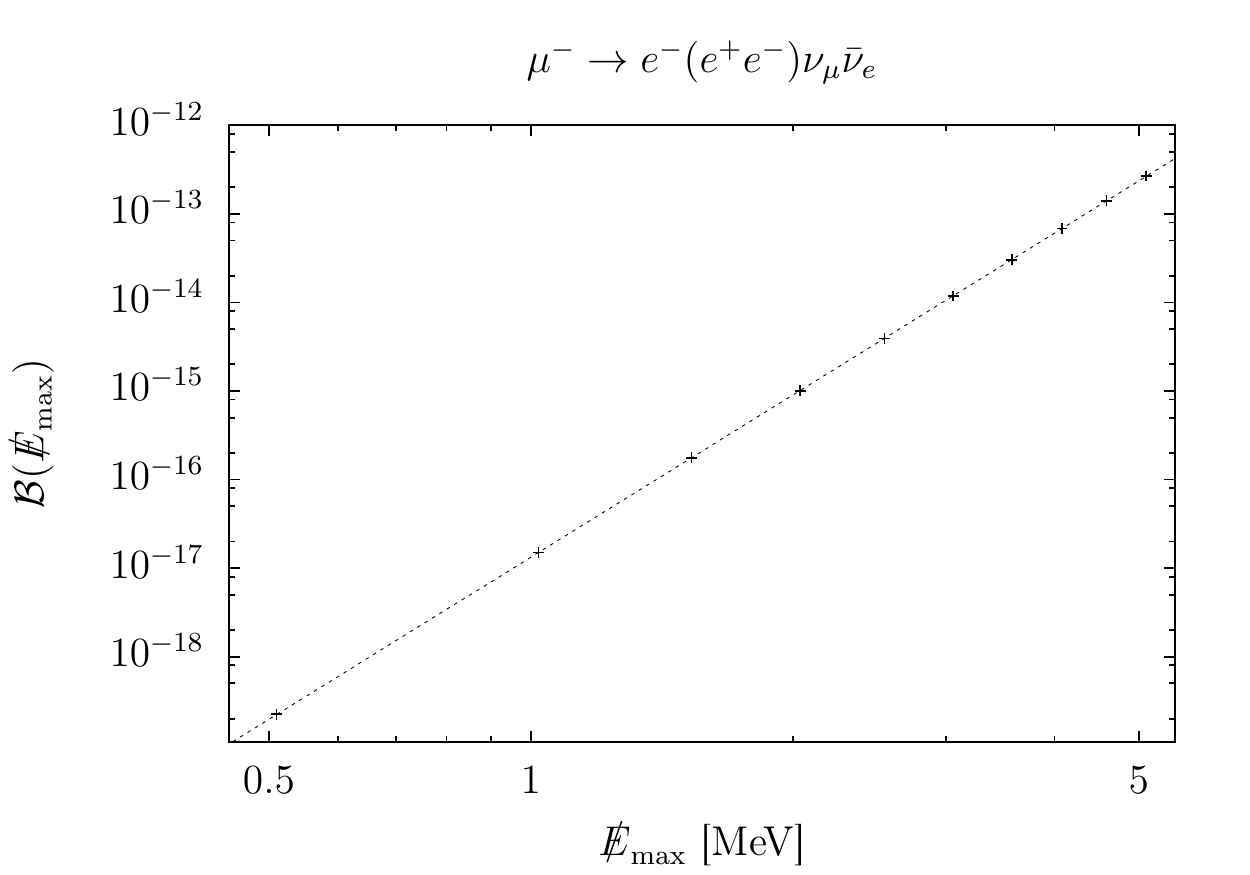}
  \caption{The branching ratios at NLO, for $\emax = 1,2,\dots,10 \, m_e$, fitted with the ansatz~\eqref{eqn:fit2}.}
  \label{fig:fit}
\end{figure}

In addition, we report in table~\ref{tab:br} the branching ratios for different missing energy cuts: $\emax = 1,5,10,20,50$ and $100 \, m_e$; our LO results are in good agreement with the values given in ref.~\cite{Djilkibaev:2008jy}.\footnote{Moreover, the NLO corrections are independently confirmed in~\cite{Pruna:2016spf}.}
In ref.~\cite{Djilkibaev:2008jy}, where the LO decay is considered, a fit of the branching ratio in the endpoint region was presented:
\begin{equation}
  \mathcal{B}(\emax) = \kappa \left( \frac{\emax}{m_e} \right)^6,
  \mbox{ with } \kappa^\text{LO}=2.99 \times 10^{-19}.
  \label{eqn:fit1}
\end{equation}
We performed a similar fit employing as input data the NLO branching ratios for $\emax =1, 2, \dots, 10 \, m_e$. Taking into account the numerical error of $\mathcal{B}(\emax)$, we obtained the following value for the constant $\kappa$ at NLO accuracy:
\begin{equation}
  \kappa^\text{NLO}=2.5117 \,(6) \times 10^{-19}.
\end{equation}
In~\eqref{eqn:fit1} the exponent of $\emax$ is fixed; relaxing such constraint and assuming it also to be a free parameter, i.e.\  
\begin{equation}
  \mathcal{B}(\emax) = \kappa' \left( \frac{\emax}{m_e} \right)^\gamma,
  \label{eqn:fit2}
\end{equation}
we obtain $ \kappa^{'\text{NLO}}=2.217\,(2) \times 10^{-19}$ and $\gamma^\text{NLO} = 6.0768 \,(4)$; our fit is also shown figure~\ref{fig:fit}.
We note that our ansatz~\eqref{eqn:fit2} is equivalent to a linear fit in the double logarithmic scale of figure~\ref{fig:fit}, $\ln \mathcal{B} = \ln \kappa' + \gamma \ln (\emax/m_e)$, while~\eqref{eqn:fit1} represents a straight line with fixed slope: $\ln \mathcal{B} = \ln \kappa + 6 \ln (\emax/m_e)$.

\section{Discussion and Conclusions}
\label{sec:con}
In this work we studied the SM prediction of the differential rates and branching ratios of the muon decay with internal conversion $\mu \to e (e^+ e^-) \nu \bar{\nu}$ at NLO. Virtual and real corrections were computed using the effective four-fermion Fermi Lagrangian plus QED and QCD, taking into account the full dependence on the mass ratio $r=m_e/m_\mu$. 

We employed the library \textsc{LoopTools} for the numerical evaluation of the coefficients appearing in the Lorenz-covariant decomposition of tensor one-loop integrals. Real corrections were calculated with a phase space slicing method. 
For photon energies below the $\omega_0$ threshold, the photon phase space integral is worked out analytically by taking advantage of the soft photon approximation for the bremsstrahlung amplitude. Above the IR cut-off, we employed the complete amplitude of the real photon emission process. 
A sufficiently small $\omega_0$ is then chosen in the calculation of the physical observables to assure in the final results the cancellation of the $\omega_0$ dependence between the soft and the hard part. Virtual and real corrections can be obtained as a Fortran code from the authors.

The branching ratio at NLO accuracy was presented in table~\ref{tab:br}. Our prediction is in agreement with the old measurement by the \textsc{Sindrum} experiment. In addition to this, we studied the decay rate as a function of the three-electron invariant mass, at the end-point region, and as a function of the cut on the missing energy.

From our results we can observe that while the global $K$-factor is close to unity --- the shift $\delta \mathcal{B}_\text{NLO}$ gives a correction of $2 \times 10^{-3}$ --- locally, in the configuration where $m_\mu - E_1-E_2-E_3 \to 0$, the relative size of radiative corrections is as large as $10-20\%$ of the LO.
Such an enhancement is given by the smallness of the $\emax$ cut, which forces the bremsstrahlung photon to be emitted in the soft-collinear region, where the corrections can behave as $(\alpha/\pi) \ln (m_e/m_\mu) \ln(\emax/m_\mu)$. 
Radiative corrections reduce the LO prediction of the width; the effect can be visualized in figure~\ref{fig:brall} shifting the curves downward. 
We estimated the theoretical uncertainty due to numerical errors in the Monte Carlo integration and renormalization scale variation. 

Our results can be employed also for the evaluation of the tau decays $\tau \to e \, (e^+e^-) \nu \bar{\nu}$ and $\tau \to \mu \, (\mu^+ \mu^-) \nu \bar{\nu}$ by properly substituting $m_\mu \to m_\tau$ and $m_e \to m_\ell$, with $\ell = e, \mu$.

\acknowledgments
In the last phase of this work we learnt about an ongoing independent calculation of the NLO corrections to $\mueee$ decay carried out by G.\ M.\ Pruna, A.\ Signer and Y.\ Ulrich.  We would like to thank them for valuable discussion and for sharing the results of their work, which are in good agreement with ours in table~\ref{tab:br}.

Moreover, we would like to thank M.\ Passera for participating in the early stages of this work. 
We are also grateful to N.\ Berger for bringing the problem to our attention and for useful discussion and correspondence.
The authors acknowledge the support from the Swiss National Science Foundation.  
\appendix

\section{Phase space decomposition}
In this appendix we discuss the phase space parametrization employed in the Monte Carlo integration of the LO, virtual and real differential widths. 

\subsection{Phase space: LO and the virtual corrections}

The generic $n$-body phase space element of a particle with momentum $P$ decaying into $n$ particles with momenta labeled by $p_1,\dots,p_n$ is
\begin{equation}
  d \Phi_n (P; p_1, \cdots, p_n) = 
  \delta^4(P - \sum_{i=1}^n p_i)
  \prod_{i=1}^n 
  \frac{d^3p_i}{(2\pi)^3 2 E_i};
  \label{eqn:PSgeneric}
\end{equation}
it can be decomposed as nested sequence of two-body pseudo-decays applying recursively the splitting formula
\begin{equation}
   d \Phi_n (P; p_1, \cdots, p_n)  =
   (2\pi)^3 dq^2 \,
   d \Phi_j (q; p_1, \cdots, p_j) \,
   d \Phi_{n-j+1} (P; q , p_{j+1}, \cdots, p_n),
    \label{eqn:splitPH}
  \end{equation}
where $q^2 = (\sum_{i=1}^j E_i)^2-\vert\sum_{i=1}^j \vec{p}_i\vert^2$.
By means of eq.~\eqref{eqn:splitPH},  we can express the five-body phase space shared by the LO and the virtual in the following way:
\begin{align}
   d \Phi_5 (P;p_1,p_2,p_3,t_1,t_2) &= 
   d \Phi_2 (t_{12}; t_1, t_2)  \,   (2\pi)^3 dt^2 \notag \\
   & \times d \Phi_2 (p_{12}; p_1, p_2)  \,   (2\pi)^3 dm_{12}^2 \notag \\
   & \times d \Phi_2 (p_{123}; p_{12}, p_3)  \,   (2\pi)^3 dm_{123}^2 \notag \\
   & \times d \Phi_2 (P; t_{12},p_{123}),
   \label{eqn:PSdecomposed}
\end{align}
where all momenta and invariant masses in~\eqref{eqn:PSdecomposed} have been defined in section~\ref{sec:lo} (see also figure~\ref{fig:LO}). 
The two-body phase space of a generic $q \to q_1 \, q_2$ decay is, in the c.m.s.,
\begin{equation}
  d \Phi_2 (q; q_1, q_2) = 
   \delta^4(q - q_1-q_2)     
   \frac{d^3q_1}{(2\pi)^3 2 E_1} 
   \frac{d^3q_2}{(2\pi)^3 2 E_2} 
   = \frac{1}{(2\pi)^6 4 M }
   \vert \vec{q}_1 \vert d \Omega_1,
   \label{eqn:PS12}
\end{equation}
where $q^2 = M^2$.
The energy and the momentum of $q_1$ and $q_2$ in their c.m.s.\ is fixed by the masses and the invariant mass of the system:
\begin{gather}
  E_1 = \frac{M^2+m_1^2-m_2^2}{2M}, \qquad
  E_2 = \frac{M^2+m_2^2-m_1^2}{2M}, \\
  \vert \vec{q}_1 \vert = \vert \vec{q}_2 \vert =
  \frac{\left[ \left( M^2-(m_1+m_2)^2 \right)
  \left( M^2-(m_1-m_2)^2 \right)\right]^{1/2}}{2M}.
\end{gather}
Employing the parametrization~\eqref{eqn:PS12} in order to express each $d\Phi_2$ in eq.~\eqref{eqn:PSdecomposed} in its own c.m.s., and performing the trivial integration over the angles $d\Omega_{123}$ and $d\phi_3^*$, we get
\begin{equation}
  d \Phi_5 (P; t_1,t_2, p_1, p_2, p_3) = 
  \frac{t \,
  \vert \vec{p}_{123} \vert \,
  \vert \vec{p_3}^{*}\vert\,
  \vert \vec{p_1}^{**} \vert}
  {512 \pi^7 m_\mu} 
  dt \, d m_{123} \, dm_{12} \,d \cos \theta_{3}^{*} \, d \Omega_{1}^{**} \, d\Phi_2 (t_{12}; t_1, t_2).
  \label{eqn:PSLOfinal}
\end{equation}
Solid angles and momenta labeled with the superscript `$*$' are in the c.m.s.\ of the two electrons and the positron, where $p_{123}^* = (m_{123},\vec{0})$, while those with `$**$' are in that one of the two electrons, where $p_{12}^{**} = (m_{12},\vec{0})$.

The allowed kinematic ranges in eq.~\eqref{eqn:integrationlimits} follow from energy conservation in each recursive splitting. Moreover, the condition on the missing energy~\eqref{eqn:emax} written in terms of the auxiliary momenta $t_{12}$ and $p_{123}$, 
\begin{equation}
  t_{12}^0 = \frac{m_\mu^2+t^2-m_{123}^2}{2m_\mu} \le \emax,
\end{equation}
leads to the set of restricted integration limits in eq.~\eqref{eqn:emaxbounds}.

\subsection{Phase space: real corrections}
The six-particle phase space element $d\Phi_6$ of the real emission process, $\mueee \gamma$, can be analogously decomposed as a series of nested two-body decays:
\begin{align}
   d \Phi_6 (P;k,p_1,p_2,p_3,t_1,t_2) &= 
   d \Phi_2 (t_{12}; t_1, t_2)  \,   (2\pi)^3 dt^2 \notag \\
   &\times  d \Phi_2 (p_{12}; p_1, p_2)  \,   (2\pi)^3 dm_{12}^2 \notag \\
   &\times  d \Phi_2 (p_{123}; p_{12}, p_3)  \,   (2\pi)^3 dm_{123}^2 \notag \\
   &\times  d \Phi_2 (q_{\nu\bar{\nu}\gamma}; k, t_{12}) \, (2\pi)^3 d q^2 \notag \\
   &\times  d \Phi_2 (P; q_{\nu\bar{\nu}\gamma},p_{123}),
   \label{eqn:PSdecomposedreal}
\end{align}
where $k$ is the momentum of the photon; we have also introduced the auxiliary momentum $q_{\nu \bar{\nu} \gamma} = k+t_1+t_2$ and its invariant mass squared, $q^2 = q_{\nu \bar{\nu} \gamma}^2$.
Let us denote with `$\ddagger$' momenta and angles in the photon-neutrinos c.m.s., where $q_{\nu\bar{\nu}\gamma}^\ddagger = (q,\vec{0})$; substituting the generic expressions for $d\Phi_2$ in eq.~\eqref{eqn:PSdecomposedreal}, and performing the trivial integration over the angles $d\Omega_{123}$ and $d\phi_\gamma^\ddagger$, we obtain:
\begin{equation}
  d \Phi_6  = 
  \frac{t \, 
    \vert \vec{k}^{\, \ddagger} \vert \,
  \vert \vec{p}_{123} \vert \,
  \vert \vec{p_3}^{*}\vert\,
  \vert \vec{p_1}^{**} \vert}
  {8 (2 \pi)^{10} m_\mu} 
  \, dq \, dt \, d m_{123} \, dm_{12} \,d \cos \theta_{\gamma}^\ddagger \, d \Omega_{3}^{*}\, d \Omega_{1}^{**} \, d\Phi_2 (t_{12}; t_1, t_2).
  \label{eqn:PSrealfinal}
\end{equation}
The integration limits given by the energy conservation in each recursive splitting are:
\begin{align}
  -1 &\le \cos \theta_{\gamma}^\ddagger \le 1,  &
  0 &\le q \le m_\mu -3 m_e, \notag \\
  -1 &\le \cos \theta_{3}^* \le 1, &
  0 & \le t \le q, \notag \\
  -1 &\le \cos \theta_{1}^{**} \le 1, &
  3 m_e &\le m_{123} \le m_\mu-q, \notag \\
  0 &\le \varphi_{3}^* < 2 \pi, &
  2 m_e &\le m_{12} \le m_{123} - m_2, \notag \\
  0 &\le \varphi_{1}^{**} < 2 \pi. &
\end{align}
To avoid the well-known IR singularity in the phase space integration, we impose the photon energy to be greater than a minimum threshold $\omega_0$:
\begin{equation}
  \omega_0 < k_0 = \gamma k_0^\ddagger 
  - \beta \gamma \cos \theta_\gamma^\ddagger \vert \vec{k}^{ \,\ddagger} \vert.
  \label{eqn:omega0condition}
\end{equation}
The Lorentz factors $\gamma$ and $\beta$, which boost the `$\ddagger$' system back into the muon rest frame, are $\beta = \vert \vec{q}_{\nu\bar{\nu}\gamma} \vert /q_{\nu\bar{\nu}\gamma}^0$ and $\gamma = q_{\nu\bar{\nu}\gamma}^0/q$.
Solving~\eqref{eqn:omega0condition} for $\cos \theta_\gamma^\ddagger$, we obtain the following restricted integration limits for $m_{123}, t$ and $\cos \theta_\gamma^\ddagger$:
\begin{gather}
  3m_e  \le m_{123} \le 
  \min \left[ m_\mu-q,\sqrt{m_\mu^2-2\omega_0 m_\mu +q^2 \left( 1-\frac{m_\mu}{2 \omega_0} \right)} \right], \notag \\
   0  \le t \le \sqrt{q^2 - \frac{2q  \omega_0  }{\gamma (1+\beta)}},\\
   -1 \le \cos \theta_\gamma^\ddagger \le 
   \min \left( 1, \frac{1}{\beta}-\frac{\omega_0}{k_0^\ddagger \gamma \beta}  \right). \notag
 \end{gather}
The constraint on the invisible energy in~\eqref{eqn:emax} --- the photon counts also as ``invisible'' in the real corrections --- can be written, in terms of the auxiliary momenta $q_{\nu \bar{\nu} \gamma}$ and $p_{123}$, as
\begin{equation}
  q_{\nu\bar{\nu}\gamma}^0 = \frac{m_\mu+q^2-m_{123}^2}{2m_\mu} \le \emax,
\end{equation}
that further reduces the integration limits of $q$ and $m_{123}$ to
\begin{gather}
  0 \le q \le \emax, \notag \\
  \sqrt{m_\mu^2-2 m_\mu \emax+q^2} \le
  m_{123} \le
  \min \left[ m_\mu-q,\sqrt{m_\mu^2-2\omega_0 m_\mu +q^2 \left( 1-\frac{m_\mu}{2 \omega_0} \right)} \right].
\end{gather}

\subsection{Phase space of the neutrinos}
The phase space of the neutrinos, $d \Phi_2 (t_{12}; t_1, t_2)$,  is left in an implicit form in eqs.~\eqref{eqn:PSLOfinal} and \eqref{eqn:PSrealfinal}. We can employ the factorization property of the matrix element in order to perform the integration analytically.
Thanks to a Fierz rearrangement of the Fermi interaction~\eqref{eqn:LFermi}, tree level, virtual and real amplitudes can be written as
\begin{equation}
  \mathcal{M} = n^\alpha \ell_\alpha,
\end{equation}
where $\ell_\alpha$ contains the spinor structure involving the muon, the positron and the electrons (also the photon in the real correction), while $n^\alpha$ contains the neutrinos' one: 
\begin{equation}
  n^\alpha = \bar{u}(t_1) \gamma^\alpha P_L v(t_2).
\end{equation}
The squared amplitude, summed over initial and final spin states, is given by
\begin{equation}
  \overline{\vert \mathcal{M}\vert^2} = 
  \sum_{\mysmall spin} n^\alpha n^{\dagger  \beta} \,
  \sum_{\mysmall spin} \ell_\alpha \ell_\beta^\dagger.
  \label{eqn:mfac}
\end{equation}
The neutrino momenta are thus enclosed only in the first factor on the r.h.s.\ of~\eqref{eqn:mfac}, while the second term depends only on their sum, $t_1+t_2$, which can be determined via momentum-energy conservation: $t_1+t_2 = P-p_1-p_2-p_3\, (-k)$.
Therefore, the integration over the neutrino phase space factorizes:
\begin{equation}
    \int d\Phi_2(t_{12}; t_1, t_2)
  \overline{\vert \mathcal{M}\vert^2}  =  
  \int d\Phi_2  (t_{12}; t_1, t_2)\sum_{\mysmall spin} n^\alpha n^{\dagger  \beta} \,
  \sum_{\mysmall spin} \ell_\alpha \ell_\beta^\dagger 
   = 
  N^{\alpha \beta} 
  \sum_{\mysmall spin} \ell_\alpha \ell_\beta^\dagger
  \label{eqn:Nexplicit}
\end{equation}
where the expression of $d\Phi_2$ can be recovered from eq.~\eqref{eqn:PSgeneric}.
By decomposing $N^{\alpha \beta}$ in a Lorentz covariant way, one can easily verify that the result is
\begin{equation}
   N^{\alpha\beta} =
   \int d\Phi_2  (t_{12}; t_1, t_2)\sum_{\mysmall spin} n^\alpha n^{\dagger  \beta} =
   \frac{1}{192 \pi^5} 
   \left( t_{12}^\alpha t_{12}^\beta - g^{\alpha\beta} t^2 \right).
 \end{equation}

\label{Bibliography}
\bibliographystyle{JHEP}
\bibliography{BIB}
\end{document}